\documentclass[reqno,11pt]{amsart}
\usepackage{hyperref}
\usepackage[mathscr]{eucal}
\usepackage{enumerate}
\numberwithin{equation}{section}

\newtheorem{Def}{Definition}[section]
\newtheorem{Thm}[Def]{Theorem}

\newtheorem{Lemma}[Def]{Lemma}

\newcommand{\beq}{\begin{equation}}
\newcommand{\eeq}{\end{equation}}
\newcommand{\Proof}{\begin{proof}}
\newcommand{\QED}{\end{proof} \noindent}

\newcommand{\R}{\mathbb{R}}

\newcommand{\nn}{\nonumber}

\allowdisplaybreaks[1]

\title[The Incompressible Relativistic Euler Equations]{A note on incompressibility of relativistic fluids and the instantaneity of their pressures}

\author[M.\ Reintjes]{Moritz Reintjes \\ \\ \today}
\address{Departamento de Matem{\'a}tica \\ Instituto Superior T{\'e}cnico \\ 1049-001 Lisbon \\ Portugal}
\email{moritzreintjes@gmail.com}
\thanks{I was funded through CAPES-Brazil as a Post-Doctorate at IMPA until December 2016. Since January 2017, I am partially supported by FCT/Portugal through (GPSEinstein) PTDC/MAT-ANA/1275/2014  and UID/MAT/04459/2013.}

\begin{document}

\begin{abstract}
We introduce a natural notion of incompressibility for fluids governed by the relativistic Euler equations on a fixed background spacetime, and show that the resulting equations reduce to the incompressible Euler equations in the classical limit as $c\rightarrow \infty$.  As our main result, we prove that the fluid pressure of solutions of these incompressible ``relativistic'' Euler equations satisfies an elliptic equation on each of the hypersurfaces orthogonal to the fluid four-velocity, which indicates infinite speed of propagation.
\end{abstract}

\maketitle

\section{Introduction}

It is intriguing to study problems of classical fluid mechanics on a relativistic level, since the character of the underlying equations may change \cite{FreistuehlerTemple}. However, as illustrated in this note, care must be taken when addressing incompressible fluids since infinite speeds of propagation may be introduced, violating the axioms of Relativity. 

The study of incompressible fluids in General Relativity began with  Karl Schwarzschild's work \cite{Schwarzschild}, addressing the special case of a static fluid. A general model analogous to incompressible fluids has been introduced by Lichnerowicz \cite{Lichnerowicz}, (see also \cite{Disconzi}), by requiring that wave speeds equal the speed of light. However, this model does not capture the rigid motion of non-relativistic incompressible fluids in the relativistic setting.

Our starting point here is to assume such rigid motion by requiring the particle number density $N$ to be constant along the fluid flow, 
\beq \label{incomp}
u^\sigma \partial_\sigma N = 0 ,
\eeq
where $u$ denotes the fluid $4$-velocity and $\partial_\sigma$ partial differentiation in some fixed coordinate system, and we always sum over repeated upper and lower indices. The relativistic Euler equations are 
\begin{eqnarray} 
\nabla_\mu (Nu^\mu) &=&0 \label{intro_Euler_1} \\ 
\nabla_\nu T^{\mu\nu} &=& 0, \label{intro_Euler_2}
\end{eqnarray}
where $T$ is the energy-momentum tensor of a perfect fluid,
\beq \label{intro_T}
T^{\mu\nu} \equiv (\rho + p) u^\mu u^\nu + p g^{\mu\nu},
\eeq
for $\rho$ the energy density and $p$ the pressure of the fluid. We assume a fixed background spacetime with metric tensor $g$ of signature $(-+++)$ and we raise and lower indices by contraction with $g$. We work in natural units ($c=1$) and assume the fluid $4$-velocity $u$ to be normalized to
\beq \label{normalization}
u^\sigma u_\sigma \equiv g_{\mu\nu} u^\mu u^\nu=-1.
\eeq
We assume that $N$, $\rho$, $u$ and $p$ are smooth (at least $C^2$) and that $N$ and $\rho$ are both strictly positive.  In our main result, Theorem \ref{Thm1}, we show that \eqref{incomp} - \eqref{intro_Euler_2} imply an elliptic equation on families of hypersurfaces, from which we expect infinite wave speeds. In light of such infinite speeds, we refer to \eqref{incomp} - \eqref{intro_Euler_2} as the incompressible \emph{covariant} Euler equations, (instead of incompressible ``relativistic'' Euler equations). 

\begin{Thm} \label{Thm1}
Assume $(N,\rho,u,p)$ is a smooth solution of the incompressible {covariant} Euler equations, \eqref{incomp} - \eqref{intro_Euler_2}, such that $\rho\equiv \rho_0$ is constant and $p+\rho>0$. Then, on each of the spacelike hypersurfaces orthogonal to $u^\mu$, the function $\Phi \equiv \ln (\rho_0 + p)$ satisfies the elliptic equation
\beq \label{intro_pressure_eqn}
\Delta_u \Phi + \dot{u}^\nu \partial_\nu \Phi  + \nabla_\nu u^\mu\: \nabla_\mu u^\nu + R_{\mu\nu} u^\mu u^\nu = 0,
\eeq
where $R_{\mu\nu}$ is the Ricci tensor of the metric, $\dot{u}^\nu \equiv  u^\mu \nabla_{\mu} u^\nu$ and $$\Delta_u \Phi \equiv \Pi^{\mu\nu} \nabla_\mu \partial_\nu \Phi$$ for $\Pi^{\mu\nu} \equiv g^{\mu\nu} + u^\mu u^\nu$. Moreover, for $\rho >0$ not constant, $p$ satisfies the elliptic equation recorded in \eqref{pressure_eqn_general} on each hypersurface orthogonal to $u^\mu$.
\end{Thm}

A key observation for the proof of Theorem \ref{Thm1} is that $\Pi^{\mu\nu}$ projects onto the spacelike hypersurfaces orthogonal to $u^\mu$ and that $\Pi^{\mu\nu}$ is a Riemannian metric on these hypersurfaces, so that $\Delta_u$ is the Laplace operator of $\Pi^{\mu\nu}$ on each of these spacelike surfaces. From the ellipticity of \eqref{intro_pressure_eqn}, we thus expect \eqref{intro_pressure_eqn} to induce infinite speed of propagation into the incompressible covariant Euler equations, violating the axioms of Relativity.\footnote{Let us clarify our expectation: On any of the spacelike hypersurfaces orthogonal to $u$, boundary values determine solutions of \eqref{intro_pressure_eqn} by standard elliptic theory \cite{Taylor}. So a signal (specifying boundary values) send from one such hypersurface to a later such hypersurface, would determine $\Phi$ instantaneously throughout that surface. This indicates that the ellipticity of \eqref{pressure_eqn} could induce infinite speed of propagation and hence instantaneity. However, infinite wave speeds could be avoided, if the elliptic equation \eqref{intro_pressure_eqn} is suitably propagated by the evolution \eqref{incomp} -  \eqref{intro_Euler_2} imposes, (analogous to the propagation of the constraint equations in General Relativity), but it is not clear to us whether such a mechanism for propagating \eqref{intro_pressure_eqn} exists.}      However, taking the classical limit of \eqref{incomp} - \eqref{intro_Euler_2}, the equations reduce correctly to the incompressible Euler equations, c.f. Theorem \ref{Thm2}. So the incompressible covariant Euler equations could be of mathematical interest, e.g., to study problems of classical fluid dynamics.

\begin{Thm} \label{Thm2}
Assume $g_{\mu\nu}$ is the Minkowski metric. Then, replacing $\rho$ in \eqref{intro_T} by $\rho c^2$ and taking the classical limit ($c\rightarrow \infty$) of \eqref{incomp} - \eqref{intro_Euler_2}, we obtain the classical incompressible Euler equations 
\begin{eqnarray} \label{incomp_Euler_classical} 
\nabla \cdot {v} = 0, \ \ \ \  \
\partial_t \rho + v\cdot \nabla \rho = 0,  \ \ \ \   \   
 \rho\left( \partial_t v + v\cdot\nabla v\right) + \nabla p = 0, 
\end{eqnarray}
where $v\in \R^3$ is the fluid velocity, $\rho$ denotes the mass density of the fluid, $\nabla$ is the gradient on $\R^3$ and $\nabla \cdot {v}$ is the divergence of $v$. Moreover, taking the classical limit of \eqref{intro_pressure_eqn} yields
\beq \label{pressure_eqn_limit}
\Delta p + \rho_0 \: \partial_\beta {v^\alpha} \,\partial_\alpha{v^\beta} =0,
\eeq
where summation is over $\alpha,\beta =1,2,3$, and $\Delta$ is the Laplace operator in $\R^3$.
\end{Thm}

Equation \eqref{pressure_eqn_limit} is the classical equation for the pressure of an incompressible fluid, c.f. \cite[(1.84)]{MajdaBertozzi}. We prove Theorem \ref{Thm1} in section \ref{Sec_instantaneity} and leave the proof of Theorem \ref{Thm2} to section \ref{Sec_Newtonian_limit}.

\section{The elliptic equation for the pressure} \label{Sec_instantaneity}

We first establish a lemma which expresses the incompressible covariant Euler equation in a form closer to \eqref{incomp_Euler_classical}.

\begin{Lemma} 
Assume $(N,\rho,u,p)$ is smooth and satisfies \eqref{incomp}. Then $(N,\rho,u,p)$ solves \eqref{intro_Euler_1} - \eqref{intro_Euler_2} if and only if $(N,\rho,u,p)$ solves 
\begin{eqnarray} 
\nabla_\sigma u^\sigma &=& 0, \label{incomp_rel_Euler_divu} \\
u^\sigma \partial_{\sigma}\rho  &=& 0, \label{incomp_rel_Euler_mass}\\
(\rho + p) u^\nu \nabla_\nu {u^\mu}  +\Pi^{\mu\nu} \partial_\nu p &=&0. \label{incomp_rel_Euler_momentum}
\end{eqnarray}
\end{Lemma}

\Proof
Our condition of incompressibility \eqref{incomp}, keeping in mind that $N$ is assumed positive, immediately implies that \eqref{intro_Euler_1} is equivalent to \eqref{incomp_rel_Euler_divu}. 

To obtain \eqref{incomp_rel_Euler_mass},  observe that the normalization of the fluid velocity \eqref{normalization} implies that
\beq  \label{1}
u^\sigma \nabla_\nu u_{\sigma}=0.
\eeq
From the expression for the energy-momentum tensor of a perfect fluid \eqref{intro_T} and from the identity \eqref{1}, we find that
\begin{eqnarray} \nn
u_\sigma\nabla_{\nu} {T^{\sigma\nu}}
&=&    u_\sigma\nabla_{\nu} \big( (\rho + p) u^\sigma u^\nu + p g^{\sigma\nu} \big) \cr
&=& (\rho + p) ( u^\nu u_\sigma \nabla_{\nu} u^\sigma  - \nabla_{\nu}  u^\nu ) - u^\nu \partial_{\nu}(\rho + p)  +  u^\sigma \partial_{\sigma} p \cr
&=& - u^\nu \partial_{\nu}\rho,
\end{eqnarray}
where we used \eqref{1} and \eqref{incomp_rel_Euler_divu} to obtain the last equality. Thus, the equation resulting from contracting \eqref{intro_Euler_2} with $u_\sigma$ is equivalent to \eqref{incomp_rel_Euler_mass}.

To obtain \eqref{incomp_rel_Euler_momentum}, observe that the normalization of the fluid $4$-velocity \eqref{normalization} implies that the projection $\Pi_{\mu\nu} \equiv g_{\mu\nu} + u_\mu u_\nu$ satisfies 
\begin{eqnarray} \label{prejector-property}
\Pi_{\sigma\mu} u^\sigma =0 \ \ \ \ \ \  \ \text{and} \ \ \ \ \ \
\Pi_{\sigma\mu} v^\sigma = v_\mu
\end{eqnarray}
for any tangent vector $v^\mu$ with $v^\sigma u_\sigma=0$. Moreover, from \eqref{1} we obtain that 
\beq \label{2}
\Pi_{\mu\sigma}  \nabla_\nu {u^\sigma} = g_{\mu\sigma}  \nabla_\nu {u^\sigma}  + u_\mu  u_\sigma\nabla_\nu  {u^\sigma}  =  \nabla_\nu u_{\mu}.
\eeq
For the energy momentum tensor of a perfect fluid \eqref{intro_T}, we find using \eqref{incomp_rel_Euler_divu}, \eqref{prejector-property} and \label{2} that
\begin{eqnarray}  \label{3}
\Pi_{\mu\sigma} \nabla_\nu {T^{\sigma\nu}} 
&=& \Pi_{\mu\sigma} \nabla_\nu \big( (\rho + p) u^\sigma u^\nu + p g^{\sigma\nu} \big) \cr
&=& (\rho+p)\Pi_{\mu\sigma} u^\nu\nabla_\nu u^\sigma +\Pi_{\mu\sigma} g^{\sigma\nu} \partial_\nu p  \cr
&=& (\rho+p)  u^\nu\nabla_\nu  u_\sigma +\Pi_{\mu\sigma} g^{\sigma\nu} \partial_\nu p.
\end{eqnarray}
Raising the index with the metric, we conclude from \eqref{3} that the equations resulting from contraction of \eqref{intro_Euler_2} with $\Pi_{\mu\sigma}$ are equivalent to \eqref{incomp_rel_Euler_momentum}.
\QED

Let us remark that \eqref{incomp_rel_Euler_divu} - \eqref{incomp_rel_Euler_mass} also follow by the standard replacement of $3$-vectors in \eqref{incomp_Euler_classical} by $4$-vectors. However, deriving \eqref{incomp_rel_Euler_divu} - \eqref{incomp_rel_Euler_mass} from \eqref{incomp}, it is clear that we captured the condition of rigid particle motion.  \vspace{.2cm} 

\noindent \emph{Proof of Theorem \ref{Thm1}:} 
Assume that $\rho\equiv \rho_0$ is constant and $p+\rho>0$.  Then, introducing $\Phi \equiv \ln (\rho_0 + p)$, we write \eqref{incomp_rel_Euler_momentum} in its equivalent form 
\beq \label{incomp_rel_Euler_momentum2}
u^\nu \nabla_\nu {u^\mu} +\Pi^{\mu\nu} \partial_\nu \Phi = 0. 
\eeq
Observe that $\Pi_{\mu\nu} \equiv g_{\mu\nu} + u_\mu u_\nu$ together with \eqref{incomp_rel_Euler_divu} imply
\beq \nn
\nabla_\mu \big( \Pi^{\mu\nu} \partial_\nu \Phi \big) = \Delta_u \Phi + \dot{u}^\nu \partial_\nu \Phi,
\eeq
where we set $\dot{u}^\nu \equiv  u^\mu \nabla_\mu u^\nu$ and $\Delta_u \Phi \equiv \Pi^{\mu\nu} \nabla_\mu \partial_\nu \Phi$. Moreover, \eqref{incomp_rel_Euler_divu} together with the definition of the Riemann curvature tensor of $g$, imply
\beq \nn
u^\nu \nabla_\mu  \nabla_\nu {u^\mu}  =  R^\mu_{\ \sigma \mu\nu} u^\sigma u^\nu = R_{\mu\nu} u^\mu u^\nu ,
\eeq
where $R_{\mu\nu}$ denotes the Ricci tensor. 
Thus taking the divergence of \eqref{incomp_rel_Euler_momentum2} gives us \eqref{intro_pressure_eqn}, that is, 
\beq \label{pressure_eqn}
\Delta_u \Phi +  \dot{u}^\nu \partial_\nu \Phi + \nabla_\nu {u^\mu} \nabla_\mu {u^\nu}  + R_{\mu\nu} u^\mu u^\nu = 0.
\eeq

We now show that \eqref{pressure_eqn} is an elliptic equation on each of the spacelike hypersurfaces orthogonal to $u$. By the second identity in \eqref{prejector-property} it follows that $\Pi_{\mu\nu}$ defines a Riemannian metric on each of these hypersurfaces. Thus $\Delta_u  \equiv \Pi^{\mu\nu} \nabla_\mu \partial_\nu$ is the Laplace operator on the surfaces orthogonal to $u$. Moreover, since $\nabla_\nu u^\mu = \Pi^{\mu\rho} \nabla_\nu u_{\rho}$, we have $\dot{u}^\nu =\Pi^\nu_{\ \rho} \dot{u}^\rho$ and it follows that the differentiation $\dot{u}^\nu \partial_\nu \Phi$ is taken tangential to the hypersurface orthogonal to the fluid flow $u$. The remaining terms in \eqref{pressure_eqn} are given source terms depending only on the metric tensor and $u$. In summary, we proved that \eqref{pressure_eqn} is an elliptic equation in $\Phi$, defined on each of the spacelike hypersurfaces orthogonal to $u$. 

To prove the supplement, observe that a computation similar to that leading to \eqref{pressure_eqn}, yields 
\beq \label{pressure_eqn_general}
\frac{1}{\rho + p} \left(\Delta_u p +  \dot{u}^\nu \partial_\nu p\right) - \frac{\Pi^{\mu\nu}}{(\rho + p)^2}\partial_{\nu}p \; \partial_\mu(\rho+p) + \nabla_\nu {u^\mu} \nabla_\mu {u^\nu}  + R_{\mu\nu} u^\mu u^\nu = 0,
\eeq
which is again an elliptic equation on each of the hypersurfaces orthogonal to the fluid $4$-velocity. This completes the proof.
\hfill $\Box$  \vspace{.2cm}

\section{The Newtonian limit} \label{Sec_Newtonian_limit}

In this section we compute the classical limit ($c\rightarrow \infty$) of the incompressible covariant Euler equations, \eqref{incomp} - \eqref{intro_Euler_2}, and of the pressure equation \eqref{intro_pressure_eqn}. We thereby prove Theorem \ref{Thm2}.  \vspace{.2cm}

\noindent \emph{Proof of Theorem \ref{Thm1}:} 
Assume $g_{\mu\nu}=\eta_{\mu\nu}$ is the Minkowski metric and assume coordinates $(x^0,...,x^3)=(ct,x,y,z)$, where $(x,y,z)$ are Cartesian coordinates on $\R^3$. To work in SI-units, we replace $\rho$ in \eqref{intro_T} by $\rho c^2$, (for $c>1$ the speed of light in SI-units), and we relate the fluid $4$-velocity $u^\mu$ to the fluid velocity $v$ by setting\footnote{By our convention $u^\mu$ is dimensionless, in contrast to the convention in \cite{Choquet,Weinberg}.} 
\beq \label{techeqn1}
u^\mu = \frac{\gamma(v)}{c} \left(\begin{array}{c} c \cr v \end{array}  \right) 
\hspace{.6cm} \text{for}  \hspace{.6cm}  
\gamma(v) \equiv \frac{1}{\sqrt{1-\frac{v^2}{c^2}} },
\eeq
which is consistent with the normalization \eqref{normalization}. From \eqref{techeqn1} we find that
\begin{eqnarray} \nn
\nabla_\mu u^\mu 
&=& \frac{\partial u^0}{\partial (ct)} + \sum_{\alpha =1}^{3} \frac{\partial u^\alpha}{\partial x^\alpha} \ = \ \frac{1}{c} \left( \frac{\partial \gamma(v)}{\partial t} + \sum_{\alpha =1}^{3} v^\alpha \frac{\partial \gamma(v)}{\partial x^\alpha} \right)  + \frac{\gamma(v)}{c} \nabla \cdot v .
\end{eqnarray}
Now, observing that 
\beq \label{techeqn2}
\frac{\partial \gamma(v)}{\partial t}= - \frac{\gamma(v)^3}{c^2}\, v\cdot \frac{\partial v}{\partial t},
\ \ \  \ \ \  \ \  \ \  
\frac{\partial \gamma(v)}{\partial x^\alpha} = -\frac{\gamma(v)^3}{c^2}\, v\cdot \frac{\partial v}{\partial x^\alpha}
\eeq
and that $\gamma(v)\rightarrow 1$ as $c\rightarrow \infty$, it follows that
\beq \nn
\lim\limits_{c\rightarrow \infty} \nabla_\mu u^\mu  = \nabla \cdot v.
\eeq
We conclude that \eqref{incomp_rel_Euler_divu} reduces to the first equation in \eqref{incomp_Euler_classical} as $c\rightarrow \infty$.

From \eqref{techeqn1}, we immediately find that
\beq \nn
u^\sigma \partial_\sigma \rho 
= \frac{\gamma(v)}{c} \left( \partial_t\rho + v\cdot \nabla \rho \right),
\eeq
from which we conclude that \eqref{incomp_rel_Euler_mass} is equivalent to the second incompressible Euler equation in \eqref{incomp_Euler_classical}.

We now compute the classical limit of \eqref{incomp_rel_Euler_momentum}. For this, introducing the notation $v^\mu\equiv \frac{c}{\gamma(v)} u^\mu$, observe that \eqref{techeqn1} implies  
 \beq \label{techeqn1b}
 u^\nu  \nabla_\nu u^\mu
= \frac{1}{c^2} \gamma(v) v^\mu v^\nu \partial_\nu \gamma(v) + \frac{1}{c^2}  \gamma(v)^2 v^\nu \partial_\nu v^\mu ,
 \eeq
from which we find that
 \beq \nn
 (\rho c^2 + p) u^\nu  \nabla_\nu u^\mu
 = \left(\rho + \frac{p}{c^2} \right)  \left( \gamma(v)  v^\nu  v^\mu \partial_\nu \gamma(v) +  \gamma(v)^2 v^\nu \partial_\nu v^\mu \right).
 \eeq
Now, since $\lim\limits_{c\rightarrow \infty} \gamma(v)=1 $ and since by \eqref{techeqn2}
 \begin{eqnarray} \label{techeqn2b}
  v^\nu   \partial_\nu\gamma(v)
 \ = \ \frac{\partial \gamma(v)}{\partial t}  + v\cdot \nabla \gamma(v) \ = \ - \frac{\gamma(v)^3}{c^2}\, v\cdot \left(  \frac{\partial v}{\partial t} + v\cdot \nabla v  \right) ,
 \end{eqnarray}
we obtain that
  \beq \nn
\lim_{c\rightarrow \infty} (\rho c^2 + p)  u^\nu \nabla_\nu {u^\mu} 
\ = \ \rho v^\nu \partial_\nu  {v^\mu}   
\ = \ \rho  \left(\begin{array}{c} 0 \cr \partial_t v + v\cdot \nabla v \end{array} \right),
 \eeq
where the $\mu=0$ component vanishes.
 Moreover, a straightforward computation yields 
 \beq \label{techeqn4}
 \Pi^{\mu\nu} = \eta^{\mu\nu} + u^\mu u^\nu \longrightarrow  \left( \begin{array}{cc} 0 & 0 \cr 0 & \text{id}_3  \end{array}  \right),
 \eeq
as $c\rightarrow \infty$, where $\text{id}_3$ denotes the identity mapping on $\R^3$. To summarize, we showed that \eqref{incomp_rel_Euler_momentum} reduces to the third equation in \eqref{incomp_Euler_classical} as $c\rightarrow \infty$.

It remains to confirm that \eqref{intro_pressure_eqn} converges to the classical pressure equation \eqref{pressure_eqn_limit} as $c\rightarrow \infty$. To begin, observe that for $g$ being the Minkowski metric we have $R_{\mu\nu}=0$ and \eqref{intro_pressure_eqn} reduces to
\beq \label{pressure_eqn_flat}
\Pi^{\mu\nu} \partial_\mu \partial_\nu \Phi + \dot{u}^\nu \partial_\nu \Phi  + \partial_\nu u^\mu\: \partial_\mu u^\nu = 0.
\eeq
Substituting $\Phi \equiv \ln(\rho_0 c^2 + p)$, we compute  for the first two terms that  
\begin{eqnarray} \nn
(\rho_0 c^2+p) \left(\Pi^{\mu\nu} \partial_\mu\partial_\nu \Phi +  \dot{u}^\nu \partial_\nu \Phi \right) 
&=& \Pi^{\mu\nu} \left( \partial_\mu\partial_\nu p - \frac{ \partial_\mu p \partial_\nu p}{\rho_0 c^2 +p} + \dot{u}^\nu \partial_\nu p \right).
\end{eqnarray}
Recalling that $\dot{u}^\nu \equiv u^\mu \nabla_{\mu} u^\nu$, \eqref{techeqn1b} together with \eqref{techeqn2b} imply $\lim\limits_{c\rightarrow \infty} \dot{u}^\nu \partial_\nu p =0$, so that in light of  \eqref{techeqn4} we conclude
\begin{eqnarray} \nn
\lim\limits_{c\rightarrow \infty}\; (\rho_0 c^2+p) \left(\Pi^{\mu\nu} \partial_\mu\partial_\nu \Phi +  \dot{u}^\nu \partial_\nu \Phi \right) \: =\:   \Delta p.
\end{eqnarray}
To handle the last term in \eqref{pressure_eqn_flat}, use  \eqref{techeqn2}, \eqref{techeqn1b} and  \eqref{techeqn2b}, to compute 
\beq \nonumber
\lim\limits_{c\rightarrow \infty}\; (\rho_0 c^2+p) \partial_\nu u^\mu \partial_\mu {u^\nu} 
\; = \; \lim\limits_{c\rightarrow \infty}\; (\rho_0 c^2+p) \frac{\gamma(v)^2}{c^2} \partial_\nu v^\mu \partial_\mu {v^\nu} 
\; = \; \rho_0  \lim\limits_{c\rightarrow \infty} \partial_\nu v^\mu \partial_\mu {v^\nu} ,
\eeq
and since $\partial_\mu v^0 = 0$,  we obtain that
\beq \nonumber
\lim\limits_{c\rightarrow \infty}\; (\rho_0 c^2+p) \partial_\nu u^\mu \partial_\mu {u^\nu} 
\; = \; \rho_0   \partial_\beta {v^\alpha} \partial_\alpha {v^\beta},
\eeq
where summation is over $\alpha,\beta =1,2,3$. In summary, we proved that multiplying \eqref{pressure_eqn_flat} by $(\rho_0 c^2+p)$, the resulting equation converges to the sought after equation \eqref{pressure_eqn_limit} as $c\rightarrow \infty$. This proves Theorem \ref{Thm2}.    \hfill $\Box$


\section*{Acknowledgments}
I thank Blake Temple for helpful discussions.

\providecommand{\bysame}{\leavevmode\hbox to3em{\hrulefill}\thinspace}
\providecommand{\MR}{\relax\ifhmode\unskip\space\fi MR }
\providecommand{\MRhref}[2]{%
  \href{http://www.ams.org/mathscinet-getitem?mr=#1}{#2} }
\providecommand{\href}[2]{#2}

\end{document}